# NO$_2$ Gas Sorption Studies of Ge$_{33}$Se$_{67}$ Films Using Quartz Crystal Microbalance

*Velichka Georgieva[1], Maria Mitkova[2,\*], Ping Chen[2], Dmitri Tenne[3], Kasandra Wolf[2] and Victoria Gadjanova[1]*

[1]Laboratory of Acoustoelectronics, Georgi Nadjakov Institute of Solid State Physics (ISSP), Bulgarian Academy of Sciences (BAS), Bulgaria

[2]Dept. of Electrical and Computer Engineering, Boise State University, Boise, ID 83725-2075, USA

[3]Dept. of Physics, Boise state University, Boise, ID 83725-1570, USA

Abstract

A study on the NO$_2$ gas sorption ability of amorphous Ge$_{33}$Se$_{67}$ coated quartz crystal microbalance (QCM) is presented. The thin films have been characterized before and after sorption/desorption processes of NO$_2$ by energy-dispersive X-ray spectroscopy (EDS), grazing angle X-ray diffraction (GAXRD), Raman spectroscopy, X-ray photoelectron spectroscopy (XPS) and atom force microscopy (AFM) measurements. These studies indicated that physisorption occurs when NO$_2$ gas molecules are introduced into the chalcogenide film and the thin film composition or structure do not change. The mass loading due to NO$_2$ gas sorption was calculated by the resonator's frequency shift. At the conditions of our experiment, up to 6.8 ng of the gas was sorbed into 200nm thick Ge$_{33}$Se$_{67}$ film at 5000 ppm NO$_2$ concentration. It has been established that the process of gas molecules sorption is reversible.

---

[\*] Corresponding author at: Dept. of Electrical and Computer Engineering, Boise State University, Boise, ID 83725-2075, USA. Tel.: +1 208 4261319; fax: +1 208 4262470.
E-mail address: mariamitkova@boisestate.edu (M. Mitkova).

I. INTRODUCTION

The importance of environmental monitoring increases with the industrialization of the planet. One of the toxic gases, subject of constant interest in this respect is $NO_2$, whose concentration grows due to combustion, increased number of vehicles in use, etc. Among the plurality of gas sensor systems that could be used for $NO_2$ monitoring, we have chosen the quartz crystal microbalance (QCM) system [1]. We value this system especially because of its unique property to detect mass at the nano-gram scale. Other attributes of such systems are an absolute measurement precision value ± 0.1 Hz, applicability in a wide concentration range – from ppm to ppb, independence of the signal from strong electric, magnetic and radiation fields, a suitable signal for digital information processing, and long-term stability performance of the elements. The thin film which comes in contact with and absorbs the gaseous molecules is a critical component for the good performance of the sensors, independently of the sensing method. In this respect the technology of inclusion of thin oxide films – $MoO_3$ and $WO_3$ [2] – is the most mature. It offers good sensitivity and response but requires additional procedures for formation of a well organized crystalline structure of the films and a high operating temperature. So, these types of films are highly preferred in cases when high operating temperatures are required, such as the automotive industry, but limitations related to their operational requirements and relatively expensive technology have stimulated a search of new media for gas sorption.

Alternative materials which have gained serious attention in this respect are chalcogenide glasses – compounds containing one or more element(s) from the sixth group of the Periodic Table (S, Se or Te). These glasses are usually created by combination of the chalcogens with elements from the fourth and fifth groups of the Periodic Table, thus forming a covalent bonding with them. The chalcogens occupy lower rows in the group VI of the Periodic Table compared to oxygen, which means that the chalcogenide compositions would be more flexible since the chemical bonds in them would be much weaker than in oxides. Due to the lack of periodicity in their structure their density is lower than that of the corresponding crystals, which contributes to their high sorption and desorption ability and high free internal volume where the adsorbed gas molecules can be accommodated.

Chalcogenide glasses have been studied in different aspects and composition combinations, the simplest among them being pure chalcogens, for example, tellurium. It has a very high molar volume of 20.46 cm$^3$/mol and thus, compared to other chalcogenide glasses [3] offers substantial potential for application in gas sensors. It shows a relatively good sorption which is governed by its two-dimensional structure that is accomplished by hexagonal chains. The experiments for improvement of its sorption and response ability are carried out in several aspects. One of them is a low temperature annealing of the films, which, however, leads to their crystallization and structural densification at temperatures over 80$^o$C, and diminishing the films sorption sensitivity [4]. Annealing of Te films to 600$^o$C leads to crystallization and formation of tubular structure which is favorable for the intensive diffusion of $NO_2$ molecules but there are concerns about a decrease of the sorption selectivity of this material and interference with other gases such as CO, $O_3$, $SO_2$, and $NH_3$ [5]. There is obvious progress in the understanding of the applicability of Te for $NO_2$ sensing, but because of its toxicity and crystallization ability we focused our studies to another chalcogenide system based on Ge-Se thin films as the absorbing medium for $NO_2$. They offer a solution to the existing problems with the two-dimensional structure of the Te chains by building a three-dimensional network instead. Experiments show that Ge stabilizes the structure, durability and stability of the pure chalcogen films towards oxidation [6]. The $Ge_{33}Se_{67}$ glasses have one of the lowest packing density among the chalcogenide glasses [3] and hence, there are chances for a good sorption ability of $NO_2$ in them.

In the present work we report our data related to sorption/desorption processes of $NO_2$ in a QCM with $Ge_{33}Se_{67}$ active film and discuss the structural prerequisites for the film's performance and the results of the films characterization before and after the gas sorption/desorption events.

II.     EXPERIMENTAL

Ge-Se bulk glass samples of typically 10 g in size were synthesized using 99.999% elemental Ge pieces and Se shots from Cerac Inc. The starting materials were weighed and then sealed in evacuated dry quartz ampoules of 10 mm internal diameter and 1 mm wall thickness. Temperature of the quartz tubes was slowly increased to 950 °C at a rate of 1 °C/min. Melts were homogenized by mixing them at 950 °C overnight. Bulk glass samples were realized by quenching melts from 50 °C above the liquidus temperature into cold water. Then Ge-Se thin films were carefully deposited by thermal evaporation onto both sides of quartz resonators. The quartz resonators were commercial AT-cut 16 MHz QCM's with 4 mm circular gold electrodes on both sides. The film thicknesses range from 50nm to 200nm and the elemental composition of as-deposited films was verified as 33.2 ± 1.1% showing good homogeneity by energy-dispersive X-ray spectroscopy (EDS) on a LEO 1430VP Scanning Electron Microscope with EDS accessory. Our preliminary results showed that the $NO_2$ gas sorption was the most prominent with the thickest $Ge_{33}Se_{67}$ film of 200 nm. Because of that, we chose to focus on samples with 200 nm thickness only for most experiments.

The films were characterized before and after the gas sorption/desorption process with grazing angle X-ray diffraction (GAXRD), Raman spectroscopy, X-ray photoelectron spectroscopy (XPS) and atomic force microscopy (AFM), in order to study their structure details and surface morphology.

Grazing XRD measurements were made on a Bruker AXS D8 Discover X-ray diffractometer with Cu K$\alpha_1$ radiation ($\lambda$=1.5406 Å) using a 1° grazing incidence geometry over 4° to 100° in 2θ (step size is 0.1° and 6 sec/step) on a NaI(Tl) scintillation detector, which improved the exact bond lengths evaluation in the structure of the films. With grazing angle setup, we were able to maximize the signal from the chalcogenide thin film layer rather than from the substrate.

Raman spectra were measured using a Horiba Jobin Yvon T64000 triple monochromator equipped with a liquid-nitrogen-cooled multichannel coupled-charge-device detector. The samples were excited with the 441.6 nm line of a He-Cd laser with the power of 60 mW focused into a circular spot of ~ 0.2 mm in diameter. The sample chamber was pumped down to $10^{-5}$ Torr range and the samples were cooled down to 100K during Raman measurements to avoid photo-induced effects due to laser irradiation. For the thin film samples, we did not see any photo darkening effects under microscope after the Raman laser irradiation and the line shapes remained the same over time. Thus we are confident that the laser power, vacuum in the samples chamber and 100K cooling conditions used for Raman experiments are sufficient for obtaining reproducible results without causing additional light induced effects. After the measurement, the Raman spectra were corrected for baseline and deconvolution was performed.

The X-ray photoelectron spectroscopy (XPS) was measured using a Physical Electronics Versaprobe spectrometer. The samples were irradiated with a monochromated Al K$_\alpha$ X-ray beam approximately 100 μm in diameter at about 100 watts scanned over a 1.4 mm x 0.1 mm area. The film samples were mounted on sample stage by securing edges with washers and screws. The spectrometer pass energy was set at 117.5 eV for the survey scan and 46.95 eV for the high resolution spectra, and the binding energy scale was calibrated using the Cu $2p_{3/2}$ and Au $4f_{7/2}$ peaks from freshly sputter cleaned 99.9% pure Cu and Au foils (Alfa Aesar). The spectrometer acceptance window was oriented for a take-off angle of 45° from the sample normal. These conditions produce full width at half maximum of better than 0.92 eV for Ag $3d_{5/2}$. To minimize sample charging, low energy electrons and Ar ions bleeding over the sample was applied.

The surface roughness and morphology of the films were studied by atomic force microscopy (AFM) with Veeco Dimension 3100 Scanning Probe Microscope with a Nanoscope V controller. This study was essential since it is known that the surface roughness and morphology are of great importance for the sorption process [1].

A schematic diagram of the gas sorption experimental setup is presented in Fig. 1. The first step was to measure the equivalent dynamic parameters of the resonators to be sure that they have kept their parameters after the process of films depositions. All samples were in a good condition and ready for the sorption experiments. Then the investigated QCM was installed on a special holder inside the test chamber. The temperature of the sample was measured by a Pt-thermosensor positioned near the QCM. First, the chamber was scavenged by 99.999% purity synthetic air from Messer, and then $NO_2$ test gas synthesized from 99.999% $N_2$ gas and 99.95% $O_2$ gas from Messer was mixed with carrier gas of synthetic air and released as a permanent flow. The velocities of both the carrier and test gases were measured and controlled by mass flow controllers and their ratio being defined by the desired concentration. The initial concentration of $NO_2$ gas is 10,000 ppm.

A frequency counter (Hameg 8123) connected to the QCM as well as to the computer for data recording registered the QCM frequency. In this way, the frequency change as a function of mass-loading during the time was identified. As an initial frequency value, we took the measured one in the carrier gas flow, under saturation conditions. The gas to be tested came from certified bottles diluted with synthetic air. The test gas was added to the carrier gas continuously for obtaining the desired composition. After adding the mixtures of gases into the system, the frequency started to decrease, and after a certain period of time it reached a constant value, when a dynamic equilibrium of certain gas concentration was established at the constant temperature in the chamber. In our experiments a temperature of 28.6$^o$C was maintained in the test camera. The $NO_2$ concentration in the gas flow was changed from 100 to 5000 ppm. At the end of the experiments, the pollutant flow was terminated and the frequency was measured to evaluate the process of desorption of mass from the QCM. At that stage, the measurement was completed.

## III. RESULTS

*X-ray diffraction.* We used a grazing angle setup to maximize the signal ratio from the chalcogenide thin film layer compared to that from the substrate. The grazing XRD results are presented in Fig. 2. The plot shows a prepeak lying at 0.88 Å$^{-1}$ smaller than $Q_p$, - the position of the principal peak of the diffraction pattern, which is determined by the nearest-neighbor distance $r_1$ in real space. This prepeak is the so-called 'first sharp diffraction peak' (FSDP) and it corresponds to real-space structural correlations on length scales appreciably larger than $r_1$, which is in the medium range order (MRO) range. The effective periodicity, $R$, can be related to the position of the FSDP, $Q_1$ [7]

$$R \approx 2\pi / Q_1 \qquad (1)$$

The correlation length, $D$, over which such quasi-periodic real-space density fluctuations are maintained, can be obtained from the full width at half maximum (FWHM), $\Delta Q_1$, of the FSDP [7] using the expression

$$D \approx 2\pi / \Delta Q_1 \qquad (2)$$

It has been proposed by Elliott that the FSDP can be represented by a structural model in which ordering of interstitial voids occurs in the structure [7]. Also Blétry [8] has given a simple formula to relate FSDP to the cation-cation nearest-neighbor distance $d$ for $AX_2$ type materials, namely,

$$d \approx 3\pi / 2Q_1 \qquad (3)$$

Based on these equations, we obtained the following data for the studied films:

$Ge_{33}Se_{67}$: $Q_1 = 0.88$ Å$^{-1}$, $\Delta Q_1 = 0.31$ Å$^{-1}$, Effective periodicity, $R \approx 2\pi / Q_1 = 7.14$ Å; Correlation lengths, $D \approx 2\pi / \Delta Q_1 = 20.3$ Å; cation-cation distance $d = 5.35$ Å

The result shows that the cation-cation distance is larger compared to $Ge_{20}Se_{80}$ thin film [3] and about 57% greater than the length of $NO_2$ molecule (3.4 Å) [9]. This suggests possibility for a relatively weak and reversible interaction of $NO_2$ molecule with the network backbone.

*Raman scattering.* Raman spectroscopy is one of the best methods to determine the structure units building the amorphous materials. Inelastic light scattering is known to be sensitive to materials' structure (the type of the structural units, their interconnection and amount), and we expected to observe some relative changes in the intensity of the vibration modes by comparing the Raman spectra before and after gas sorption. Raman spectra of $Ge_{33}Se_{67}$ thin films before and after $NO_2$ gas sorption are plotted in Fig. 3. Deconvolution of the measured Raman spectra was performed in order to distinguish the vibration modes having contribution in the integrated light scattering from each sample. In the spectra of $Ge_{33}Se_{67}$ films, the deconvolution distinguished 4 bands located at 175, 200, 280 and 310 cm$^{-1}$. Based on the commonly accepted interpretation [10-13], we assign peak at 175 cm$^{-1}$ to symmetric stretching mode of ethane-like Ge-Ge homopolar bond structures (ETH: (Se$_{1/2}$)$_3$-Ge-Ge-(Se$_{1/2}$)$_3$ units) [10, 14]. The band around 200 cm$^{-1}$ has originated from the symmetric stretching mode of corner-sharing (CS) and edge-sharing (ES) GeSe$_4$ tetrahedral structural units. The broad bands in the range 270–320 cm$^{-1}$ are due to vibration modes of Se chains and asymmetric vibrations of ETH/CS/ES structural units. Compared to the Raman spectra of bulk glass with the same composition, the results reveal much stronger vibrations from Ge-Ge and Se-Se homopolar bonds in the thin films. We suggest that the increased number of wrong bonds is related to the fact that the thin films usually contain more defects due to the increased free surface to volume ratio and hence higher number of dangling bonds. The spectra did not show much change before and after NO$_2$ gas sorption, indicating that the structure is not changed by NO$_2$ molecules and the gas sorption process is related to physisorption.

*X-ray photoelectron spectroscopy.* Fig. 4 shows Ge 3d core level XPS spectra of as-deposited Ge$_{33}$Se$_{67}$ film and the same batch film after NO$_2$ gas sorption. Each core-level spectrum was fitted with two sets of spin-orbit-split doublets corresponding to **Ge**-Se and **Ge**-O bonds respectively. Concerning the Ge 3d peak fitting, two parameters specific of the doublet were fixed for each contribution: (i) the area ratio (2*5/2+1):(2*3/2+1) = 1.5 and (ii) the energy difference between the 3d$_{5/2}$ and 3d$_{3/2}$ peak positions is 0.56 eV. These reference values were obtained experimentally from standards. Also, the full-width at half-maximum (FWHM) was assumed to be the same for the peaks within one doublet but difference between FWHM values for different doublets was allowed. The ratios of the contribution from **Ge**-Se or **Ge**-O bonds are labeled in the figure. In Ge 3d core level XPS spectra, unlike the previously reported results in [3], we saw a slight decrease of **Ge**-O bonds after gas sorption. This suggests that the film was not subject to additional oxidation by $NO_2$ gas and gas sorption is pure physical.

*Atomic force microscopy.* It is known that the performance of the gas sensor greatly depends on the surface morphology and surface roughness change could also be an indicator of trapping of $NO_2$ molecules. Measurements were performed in order to get information about the surface morphology of the films. It can give an idea about the sorption ability of the films since this is related to the structure of the material and the free surface which is bigger at rougher structures. There were AFM data collected from different points of all samples showing that the films are relatively smooth. The AFM scans for 200 nm thick film are shown in Fig. 5. The surface roughness $R_q$ of virgin samples is 2.05 nm while after gas sorption the roughness increased slightly to 2.16 nm.

*$NO_2$ gas sorption and desorption properties.* Quartz crystal microbalance is capable of measuring nano-gram scale weight changes due to $NO_2$ gas sorption or desorption. By applying the Sauerbrey equation [15] which is related to AT-cut quartz resonators, we were able to correlate the sorbed mass $\Delta m$ during each experiment with the frequency change $\Delta f$

$$\Delta f = -2.26 \cdot 10^6 f_0^2 \Delta m / s \qquad (4)$$

where $\Delta f$ and $f_0$ are in Hz and MHz, $\Delta m$ is in gram and $s$ in $cm^2$ is the area of the QCM electrode.

Fig. 6. shows the time-frequency characteristic of a 200nm thick $Ge_{33}Se_{67}$ thin film sample at a constant concentration of the $NO_2$ of 1000 ppm. The measurement started when the frequency of the QCM-$Ge_{33}Se_{67}$ system in the synthetic air flow became constant. In the $NO_2$

flow the system was constantly powered on for resonance frequency measurements. The program completed frequency measurement for each data point in a very short time interval of 2 seconds and the frequency resolution is limited to 1 Hz but can be improved to 0.2 Hz with longer measurement time. Because of these experimental conditions the graph has a staircase character. The measured frequency decreased linearly during gas sorption process and saturated at approximately 20 minutes. The total decrease in the frequency between the initial frequency and the saturated one is 10 Hz and corresponds to a weight gain of 2.3 ng due to $NO_2$ gas sorption. Then the synthetic air was forced through the system for 10 minutes and the sorbed $NO_2$ gas started desorption, i.e. the process is *reversible* at the purging conditions. The measured frequency showed good linearity with desorption time. Based on extrapolation of the desorption curve, the estimated full desorption time is 30 minutes and the desorption rate is approximately 1.5 times of the sorption rate.

We also checked how the structure responds as a function of the $NO_2$ concentration. When the $NO_2$ gas phase concentration increases, there is additional mass loading when the saturation under the particular concentration step is reached. The process is step like as shown in Fig. 7 and the weight gain and gas sorption rate during each step are summarized in Table 1. The sorption rate is almost constant around 0.1 ng/min at all $NO_2$ concentration conditions from 100-5000 ppm and the sorption process is linear which is typical for physisorption.

The total weight gain at saturation for different $NO_2$ concentrations is shown in Fig. 8. As illustrated, the frequency shift and sorbed mass increase with increased $NO_2$ concentrations. For 200nm thick film, the maximum weight increase is 6.8 ng at 5000 ppm $NO_2$ concentration which corresponds to $1.48 \times 10^{-10}$ mole of $NO_2$ molecules. Since we know the molar volume of $Ge_{33}Se_{67}$ bulk glass is 17.67 cm$^3$/mol [16], we can calculate that there is a total amount of $1.42 \times 10^{-7}$ mole of Ge and Se atoms in the 200nm thick film. That gives us one sorbed $NO_2$ molecule in approximately every 962 Ge and Se atoms.

IV. DISCUSSIONS

The selectivity of the chalcogenide glasses for $NO_2$ detection has already been regarded earlier [17]. For a fast and effective sorption of $NO_2$, a large free volume and a very open structure are desirable. One initial idea about the ability of the studied glasses to sorb gas molecules can come from their molar volume [18] and packing fraction [3] presented in Fig. 9. The free-volume in the Ge-Se system is around 10-15 %, with the biggest free-volume occurring at 33.3% Ge concentration [18], i.e. at the composition of $Ge_{33}Se_{67}$ considered in the present work. This composition corresponds to a minimum in the packing fraction, as shown in Fig. 9. That means that a significant amount of volume is empty, and this empty space may aid in the migration of gaseous molecules through the material. The relatively big voids will not interact with absorbed $NO_2$ molecules as strongly as previously reported in $Ge_{20}Se_{80}$ glass [3] where the molar volume is much smaller and very tight contact between the $NO_2$ molecules and the Ge-Se backbone occurs. Our studies of the sorption effects of $NO_2$ in the Ge-Se system suggest a strong compositional dependence on this process which is in essence related to the structure of the glass. It is for this reason that in some cases the process is purely physical as shown in the present work and for example in Ref. [17], or there is a chemical component as referred to in Ref. [3].

The GAXRD study also showed that the cation-cation distance is larger in $Ge_{33}Se_{67}$ than $Ge_{20}Se_{80}$ film suggesting bigger voids in the active layer and less interaction with sorbed $NO_2$ molecules. Further structural and surface morphology characterization by Raman, XPS and AFM confirmed that there was little or no change in the chalcogenide film. Thus the $NO_2$ physisorption process is completely reversible.

Application of the $Ge_{33}Se_{67}$ films has though another specific – the initial films have a very high oxidized component as revealed by the XPS studies, whose relative participation in the spectra slightly decreases after the gas sorption process. We assume that the chemically bonded oxygen resides on the surface of the films as suggested by our earlier studies [19, 20], which is a result of Ge oxidation occurring in an oxygen-containing atmosphere. The length of the Ge-O bond is 0.19 Å and the size of the Ge-O 6-member ring is around 0.36 Å. In other words, the free channels for diffusion of the $NO_2$ molecules are much more restricted in size than in the case of pure non-oxidized structure. This implies that the diffusion process through the oxide film would be obstructed. We suggest this to be the reason for a relatively slow performance of the $Ge_{33}Se_{67}$ based devices. There is a decrease of the relative presence of the Ge-O bonds after gas absorption. In the same time we observed increase of the surface roughness of the films at this condition and we suggest that there is some interaction between the gas flow and the surface structure of the films leading to partial reorganization of the structure.

The results of this study suggest that when dealing with chalcogenide glasses with a high concentration of Ge in which a presence of chalcogenide chains is not expected, oxidation of the Ge-Se backbone could be a concern.

**CONCLUSIONS**

The $Ge_{33}Se_{67}$ thin films show relatively good physisorption sensitivity towards $NO_2$ and the process is reversible. Unlike $Ge_{20}Se_{80}$ film studied in our previous work [3], the $Ge_{33}Se_{67}$ thin film has a larger free volume based on molar volume data and GAXRD results and does not chemically interact with $NO_2$ molecules. Raman, XPS and AFM results before and after gas absorption also support the conclusion that there is no structural change after gas absorption. Thus the $NO_2$ absorption process in the studied glass films is completely reversible. However, oxidation of the Ge-Se backbone is of some concern since the formed oxide surface film could decrease the $NO_2$ diffusion rate into the Ge-Se backbone.

**Acknowledgements**

This work has been conducted as a collaboration between Boise State University and Bulgarian Academy of Sciences - Institute of Solid State Physics. The authors are thankful to the National Science Foundation (NSF) via IMI (International Materials Institute for New Functionality in Glass), Grant DMR-0844014 and DMR-1006136 for travel support and Raman instrumentation during the project. We would like to acknowledge Chad Watson, Brian Jaques and Gordon Alanko for their help with AFM, XRD and XPS measurements, respectively. AFM, XRD and XPS instruments have been supported by NSF-Major Research Instrumentation awards No. 0216312, 0619795 and 0722699.

**Figure captions**

**Figure 1.** The gas absorption experimental setup with the following basic modules: a gas module (GM) – bottles with carrier gas, purge gas and test gas; a gas mix and control module (GMCM) – which included two mass flow controllers (FC-260 and FC-280) and a mixing camera; a test chamber (TC) with a Pt-thermosensor (PS) and mass sensitive sensor (MS); a thermostat module (TM); a generator and frequency counter (GFC) and a computer system (CS).

**Figure 2.** XRD data for the $Ge_{33}Se_{67}$ films, the XRD intensity I(Q) was plotted against scattering vector Q (= $4\pi\sin\theta/\lambda$).

**Figure 3.** Deconvoluted Raman spectra of 200nm-thick $Ge_{33}Se_{67}$ glass films before and after gas sorption.

**Figure 4.** Ge 3d core level XPS spectra for $Ge_{33}Se_{67}$ thin film (200nm thickness, resonator 5-38). a) as-deposited; b) after $NO_2$ gas sorption.

**Figure 5.** AFM data for $Ge_{33}Se_{67}$ thin film. a) as-deposited; b) after $NO_2$ gas sorption -200nm thickness.

**Figure 6.** Frequency – time characteristic (FTC) of $Ge_{33}Se_{67}$-QCM at 1000 ppm $NO_2$ concentration. Thickness of $Ge_{33}Se_{67}$ – 204nm

**Figure 7.** Frequency – time characteristic (FTC) of $Ge_{33}Se_{67}$-QCM system at different $NO_2$ concentrations. Thickness of $Ge_{33}Se_{67}$ – 204nm

**Figure 8.** Dependence of the $Ge_{33}Se_{67}$-QCM frequency shift and sorbed mass towards $NO_2$ concentrations. Thickness of $Ge_{33}Se_{67}$ – 204nm

**Figure 9.** Molar volume and packing fraction of Ge-Se glasses (molar volume data taken from Ref. [18] and packing fraction data calculated according to Ref. [3]).

Table 1. Weight gain and sorption rate at different $NO_2$ gas concentration

| $NO_2$ gas concentration (ppm) | Total weight gain at saturation $\Delta m$ (ng) | Additional weight gain (ng) | Gas sorption rate (ng/min) |
|---|---|---|---|
| 100 | 0.45 | 0.45 | 0.091 |
| 500 | 1.36 | 0.91 | 0.083 |
| 1000 | 2.95 | 1.59 | 0.096 |
| 2500 | 4.76 | 1.81 | 0.109 |
| 5000 | 6.80 | 2.04 | 0.115 |

**Figure 1.**

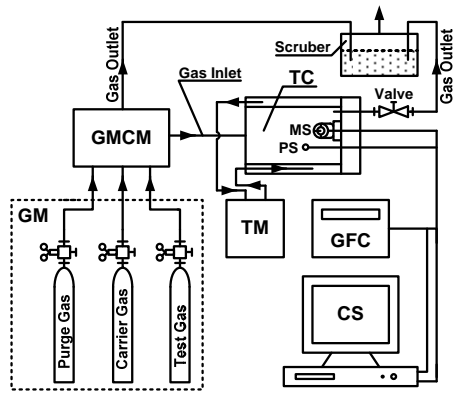

**Figure 2.**

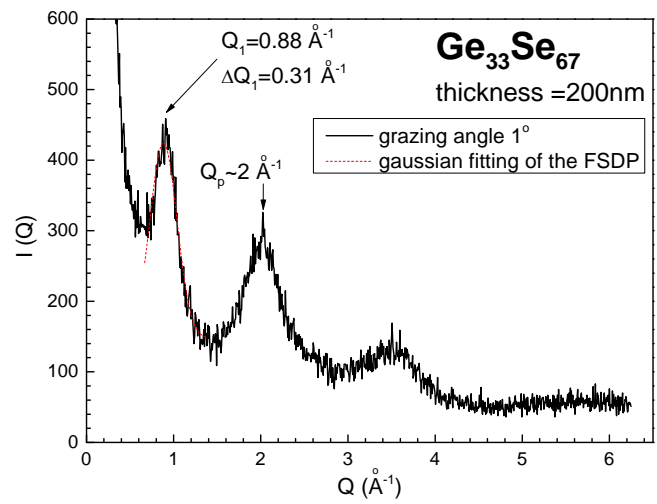

**Figure 3.**

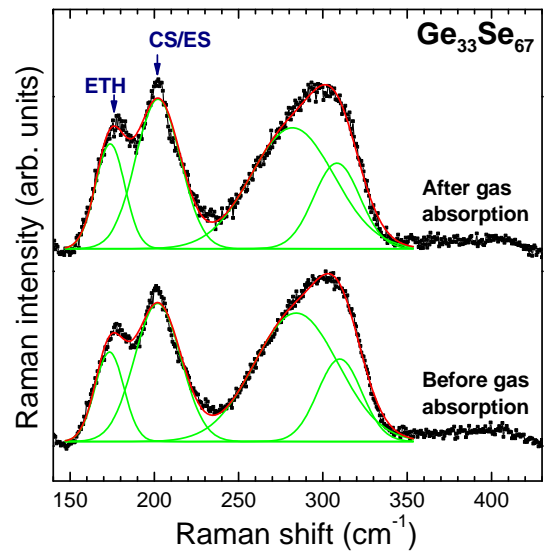

**Figure 4.**

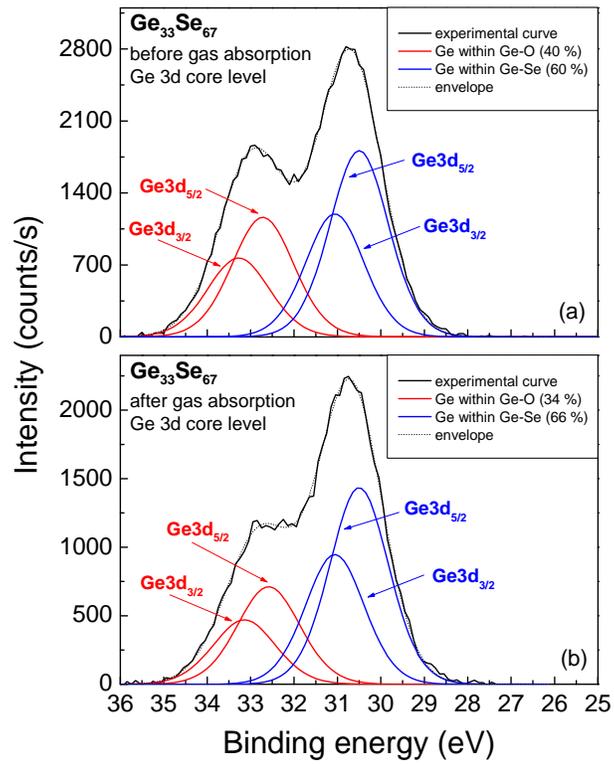

**Figure 5.**

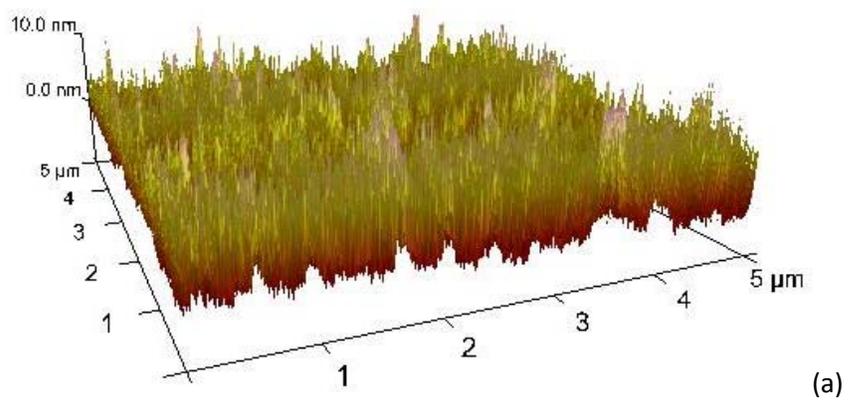

(a)

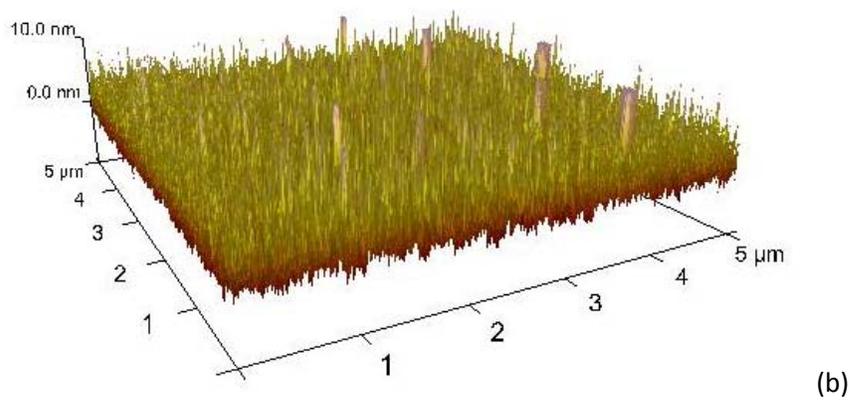

(b)

**Figure 6.**

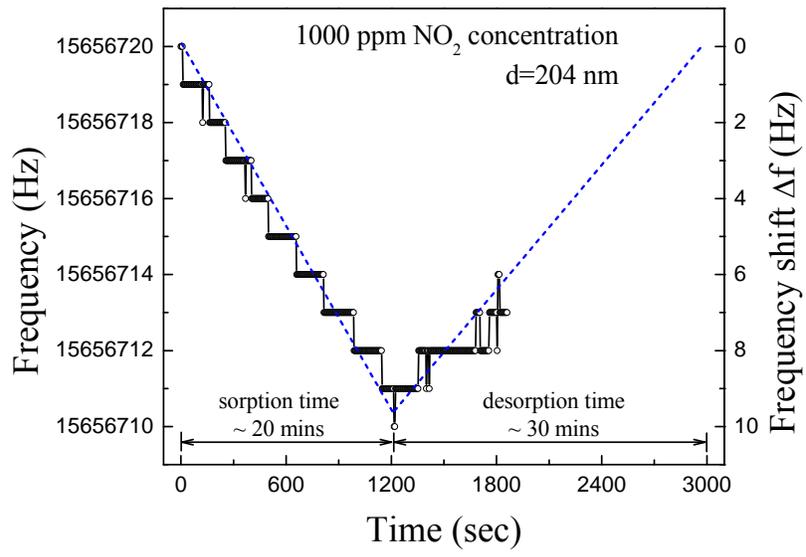

**Figure 7.**

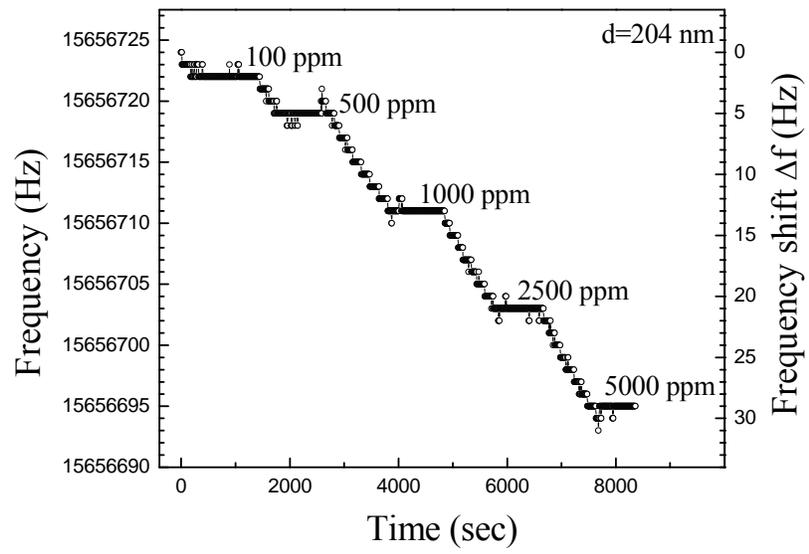

**Figure 8.**

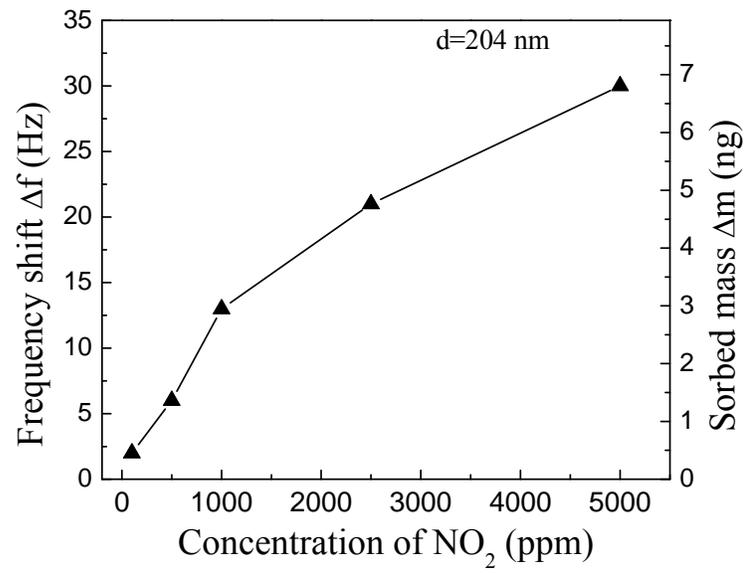

**Figure 9.**

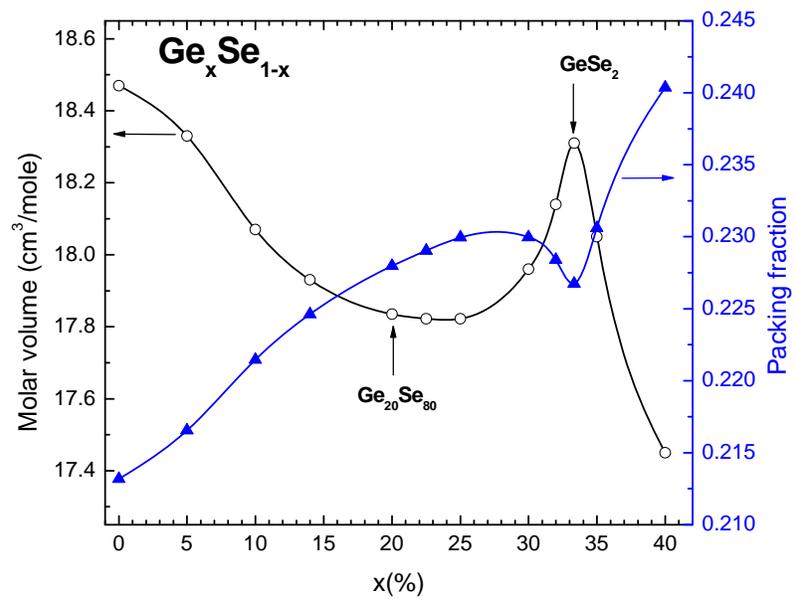